\documentclass[12pt]{article} 

\usepackage{amsmath,amssymb,amsfonts}
\usepackage{psfrag}
\usepackage{color}
\definecolor{darkblue}{rgb}{0.1,0.1,.7}
\usepackage[colorlinks, linkcolor=darkblue, citecolor=darkblue, urlcolor=darkblue, linktocpage]{hyperref} 
\usepackage[square, comma, compress,numbers]{natbib}
\usepackage[]{amsmath}
\usepackage[]{graphicx}
\usepackage[]{latexsym}
\usepackage{geometry}
\usepackage{amscd}
\usepackage[all,cmtip]{xy}
\usepackage{mathrsfs}
\usepackage[margin=10pt,font=small,labelfont=bf]{caption}
\usepackage{dsdshorthand}
\usepackage{simplewick}
\usepackage{changepage}
\numberwithin{equation}{section}

\renewcommand{\be}{\begin{eqnarray}}
\renewcommand{\ee}{\end{eqnarray}}
\newcommand{\bea}{\begin{eqnarray}}
\newcommand{\eea}{\end{eqnarray}}

\def\beq{\begin{equation}} 
\def\eeq{\end{equation}} 
\def\<{\langle}
\def\>{\rangle}

\def\nn{\nonumber} 
 
\def\cO {{\cal O}} 
 
\def\cF {{\cal F}}

\begin{document}

\vspace*{-.6in} \thispagestyle{empty}
\begin{flushright}
\end{flushright}
\vspace{.2in} {\Large
\begin{center}
{\bf Bootstrapping the $O(N)$ Vector Models \\\vspace{.1in}}
\end{center}
}
\vspace{.2in}
\begin{center}
{\bf 
Filip Kos$^{a}$, 
David Poland$^{a}$,
David Simmons-Duffin$^{b}$} 
\\
\vspace{.2in} 
$^a$ {\it  Department of Physics, Yale University, New Haven, CT 06520}\\
$^b$ {\it School of Natural Sciences, Institute for Advanced Study, Princeton, New Jersey 08540}
\end{center}

\vspace{.2in}

\begin{abstract}
We study the conformal bootstrap for 3D CFTs with $O(N)$ global symmetry. We obtain rigorous upper bounds on the scaling dimensions of the first $O(N)$ singlet and symmetric tensor operators appearing in the $\phi_i \times \phi_j$ OPE, where $\phi_i$ is a fundamental of $O(N)$. Comparing these bounds to previous determinations of critical exponents in the $O(N)$ vector models, we find strong numerical evidence that the $O(N)$ vector models saturate the bootstrap constraints at all values of $N$. We also compute general lower bounds on the central charge, giving numerical predictions for the values realized in the $O(N)$ vector models. We compare our predictions to previous computations in the $1/N$ expansion, finding precise agreement at large values of $N$.

\end{abstract}

\newpage

\tableofcontents

\newpage


\section{Introduction}
\label{sec:intro}

Conformal field theories (CFTs) offer delightful examples of quantum field theories that are strongly coupled, yet contain enough symmetry and structure that they may turn out to be tractable if the right techniques are found. Until recently, the idea of exploiting this structure in order to find complete non-perturbative solutions to theories was only carried out successfully in 2D, most notably in the seminal work of~\cite{Belavin:1984vu}. However, over the last several years great progress has been made at developing the conformal bootstrap~\cite{Ferrara:1973yt,Polyakov:1974gs} approach to CFTs in $D>2$, where a large number of nontrivial bounds have been found which follow very generally from the constraints of crossing symmetry and unitarity~\cite{Rattazzi:2008pe,Rychkov:2009ij,Caracciolo:2009bx,Poland:2010wg,Rattazzi:2010gj,Rattazzi:2010yc,Vichi:2011ux,Poland:2011ey,Rychkov:2011et,ElShowk:2012ht,Liendo:2012hy,ElShowk:2012hu,Beem:2013qxa}. The results obtained so far have been particularly striking in 3D, where it was found in~\cite{ElShowk:2012ht} that the CFT described by the critical 3D Ising model occupies a special place in the space allowed by crossing symmetry and unitarity. Moreover it appears possible that a robust numerical solution to this theory can be obtained using bootstrap techniques~\cite{ElShowk:future}.

In this paper we will extend the work of~\cite{ElShowk:2012ht} to study 3D CFTs with an $O(N)$ global symmetry using the conformal bootstrap. We will focus on theories containing a scalar field $\phi_i(x)$ in the vector representation of $O(N)$. The most notable theories falling into this class are the critical $O(N)$ vector models~\cite{Brezin:1972fb,Wilson:1973jj}, which describe second-order phase transitions in a variety of real-world systems at small values of $N$~\cite{Pelissetto:2000ek}, and are also solvable in a $1/N$ expansion at large values of $N$ (see~\cite{Moshe:2003xn} for a review). Moreover, the $O(N)$-singlet sector of this theory is thought to be holographically described by a higher-spin gauge theory in $AdS_4$~\cite{Klebanov:2002ja}. Previously, bootstrap ideas have been applied to the $O(N)$ vector models in the $1/N$ expansion, for example in work by Lang and R\"{u}hl~\cite{Lang:1990ni,Lang:1990re,Lang:1991kp,Lang:1991jm,Lang:1992zw,Lang:1993ct}, Petkou~\cite{Petkou:1994ad,Petkou:1995vu}, and more recently Maldacena and Zhiboedov~\cite{Maldacena:2011jn, Maldacena:2012sf}. Our approach allows bootstrap constraints to be studied (albeit numerically) at any value of $N$.

Our primary goal, following previous numerical studies of the bootstrap~\cite{Rattazzi:2008pe,Rychkov:2009ij,Caracciolo:2009bx,Poland:2010wg,Rattazzi:2010gj,Rattazzi:2010yc,Vichi:2011ux,Poland:2011ey,Rychkov:2011et,ElShowk:2012ht,Liendo:2012hy,ElShowk:2012hu,Beem:2013qxa}, will be to place general upper bounds on the scaling dimensions of the first nontrivial scalar operators (both $O(N)$ singlets $S$ and $O(N)$ symmetric tensors $T_{ij}$) in the $\phi_i \times \phi_j$ operator product expansion (OPE). We will also place general lower bounds on the central charge $c$, defined as the coefficient appearing in the two-point function of the stress-energy tensor.  We will then compare these bounds to the best previous results based on Monte Carlo simulations, analytical estimates, and the $1/N$ expansion. In all cases where results are known (including small values of $N$), we find that the $O(N)$ vector models saturate our bounds, and moreover sit near special locations in the space allowed by crossing symmetry. Inputting the previously measured values of $\Delta_{\phi}$, this allows us to give sharp predictions for $\Delta_{S}$, $\Delta_T$, and $c$ for different values of $N$.

In order to efficiently implement the bootstrap in the presence of a global symmetry, we will use techniques based on semidefinite programming, developed for 4D CFTs in~\cite{Poland:2011ey}. Here we will show how to adapt this technique for CFTs in an arbitrary number of space-time dimensions. This requires approximating conformal blocks as rational functions of the exchanged operator dimension $\Delta$, and we will show that such a rational approximation follows directly from a recursion relation expressing a general conformal block as a sum over poles $\sim 1/(\Delta-\Delta_*)$ occurring at special (non-unitary) values of the dimension $\Delta_*$ where the conformal multiplet contains a null state. This conformal block representation generalizes an idea of Zamolodchikov, first applied to Virasoro blocks in 2D~\cite{Zamolodchikov:1985ie,Zamolodchikov:1987}, to arbitrary space-time dimensions.

This paper is organized as follows. In section~\ref{sec:bootstrap} we review the formulation of the conformal bootstrap for CFTs containing an $O(N)$ global symmetry, as well as convex optimization techniques for placing bounds on operator dimensions and OPE coefficients. In section~\ref{sec:rational} we show how to find rational representations for conformal blocks in arbitrary space-time dimensions, presenting a new recursion relation for conformal blocks as a sum over poles in $\Delta$. In section~\ref{sec:results} we present our bounds and a comparison with the $O(N)$ vector models. We conclude in section~\ref{sec:discussion}.

\section{Conformal Bootstrap with $O(N)$ Global Symmetry}
\label{sec:bootstrap}

\subsection{Statement of Crossing Symmetry}

Let us briefly review the formulation of the conformal bootstrap for 3D CFTs with an $O(N)$ global symmetry. Further details can be found in~\cite{Rattazzi:2010yc,Vichi:2011ux,Poland:2011ey}. We focus on theories containing a scalar primary operator $\phi_i$ of dimension $\Delta_{\phi}$, transforming as a fundamental under $O(N)$.  The operator product of $\phi_i$ with itself takes the schematic form
\be
\phi_i \times \phi_j \sim \sum_{S^+} \delta_{ij} \cO + \sum_{T^+} \cO_{(ij)} + \sum_{A^-} \cO_{[ij]},
\ee
where $S^+$ denotes $O(N)$ singlets of even spin, $T^+$ denotes $O(N)$ symmetric tensors of even spin, and $A^-$ denotes $O(N)$ anti-symmetric tensors of odd spin.

By pairing up $\f$'s and performing the OPE, a four-point function can be decomposed into conformal blocks as
\bea
x_{12}^{2\Delta_{\phi}} x_{34}^{2\Delta_{\phi}}
\<
\contraction{}{\f_i(x_1)}{}{\f_j(x_2)}
\f_i(x_1)\f_j(x_2)
\contraction{}{\f_k(x_3)}{}{\f_l(x_4)}
\f_k(x_3)\f_l(x_4)
\>
 &=& \sum_{S^+} \lambda^2_{\cO} (\delta_{ij} \delta_{kl}) g_{\Delta,\ell}(u,v) \nn\\
&& + \sum_{T^+} \lambda_{\cO}^2 \left( \delta_{il} \delta_{jk} + \delta_{ik}\delta_{jl} - \frac{2}{N} \delta_{ij} \delta_{kl} \right) g_{\Delta,\ell}(u,v) \nn\\
&& + \sum_{A^-} \lambda_{\cO}^2 (\delta_{il}\delta_{jk} - \delta_{ik}\delta_{jl} ) g_{\Delta,\ell}(u,v),
\label{eq:vectorOPE}
\eea
where each sum runs over {\it primary} operators $\cO$ of dimension $\De$ and spin $\ell$ appearing in $\f_i\x \f_j$.  Here, 
 $x_{ij} \equiv x_i - x_j$, $\lambda_{\cO}$ is the OPE coefficient of $\cO$, and the conformal blocks $g_{\De,\ell}(u,v)$ are functions of conformal cross-ratios 
\be
u = z \bar{z} = \frac{x_{12}^2 x_{34}^2}{x_{13}^2 x_{24}^2},\qquad 
v = (1-z)(1-\bar{z}) = \frac{x_{14}^2 x_{23}^2}{x_{13}^2 x_{24}^2}.
\ee

The four-point function itself should be independent of how we perform the OPE.  Swapping $(1,i)\leftrightarrow(3,k)$, we find two different conformal block expansions of a single four-point function which must agree with each other.  Writing out this condition and isolating the coefficient of each tensor structure that appears, we obtain three equations which can be grouped into a vector ``sum rule"
\newcommand\cV{\mathcal{V}}
\be
\label{eq:vectorsumrule}
\sum_{S^+}\l_\cO^2 V_{S,\De,\ell}+\sum_{T^+}\l_\cO^2 V_{T,\De,\ell}+\sum_{A^-}\l_\cO^2 V_{A,\De,\ell} &=& 0,
\ee
where
\begin{align}
\label{eq:vectordefinitions}
V_{S,\De,\ell} &= \p{
\begin{array}{c}
0 \\
F^-_{\De,\ell}\\
F^+_{\De,\ell}
\end{array}
}
,\quad
V_{T,\De,\ell} = \p{
\begin{array}{c}
F^-_{\De,\ell}\\
(1-\frac 2 N)F^-_{\De,\ell}\\
-(1+\frac 2 N)F^+_{\De,\ell}
\end{array}
}
,\quad
V_{A,\De,\ell} = \p{
\begin{array}{c}
-F^-_{\De,\ell}\\
F^-_{\De,\ell}\\
-F^+_{\De,\ell}
\end{array}
},\\
F^{\pm}_{\De,\ell}(u,v) &\equiv v^{\De_\f} g_{\De,\ell}(u,v)\pm u^{\De_\f} g_{\De,\ell}(v,u).
\end{align}
The coefficients $\l_\cO^2$ appearing in~(\ref{eq:vectorsumrule}) are unknown a-priori, with the exception of the unit operator which has $\l_\cO=1$ if $\f_i$ is canonically normalized.  However, we do know that the OPE coefficients $\l_\cO$ must be real in unitary theories, which means that $\l_\cO^2$ is positive. Further, in $D$-dimensional unitary theories the operator dimensions must satisfy the lower bounds~\cite{Ferrara:1974pt,Mack:1975je,Metsaev:1995re,Minwalla:1997ka,Grinstein:2008qk}
\begin{equation}
\label{eq:unitaritybound}
\Delta \geq 
\begin{cases}
\frac{D-2}{2}  & \text{if}\qquad \ell = 0, \\
\ell+D-2 & \text{if}\qquad \ell > 0,
\end{cases}
\end{equation}
where saturation occurs for free scalars ($\ell=0$) or conserved currents ($\ell>0$).

\subsection{Bounds from Convex Optimization}
\label{sec:convex}

From here, we follow the general strategy of \cite{Rattazzi:2008pe} for putting bounds on CFT data.  Let us begin by isolating the unit operator in~(\ref{eq:vectorsumrule}),
\be
\label{eq:unitopisolated}
0 &=& V_{\mathrm{unit}} + \sum \l_\cO^2 V_{\cO}.
\ee
The procedure is as follows:
\begin{enumerate}
\item Make an assumption about the CFT spectrum, for instance that all singlet scalars have dimension above some $\De_*$.
\item Try to find a linear functional $\a$ such that
\be
\a(V_\mathrm{unit}) &>& 0 ,\nn\\
\a(V_\cO) &\geq& 0 \textrm{ for all $\cO$ satisfying the assumption in (1)} .
\ee
\item If such a functional exists, the assumption (1) is ruled out, since applying $\a$ to Eq.~(\ref{eq:unitopisolated}) gives a contradiction.  If not, we cannot conclude anything about our assumption.
\end{enumerate}

Step 2 requires us to solve a convex optimization problem: we must search over the vector space $\cF$ of linear functionals, subject to linear constraints of the form $\a(V)\geq 0$.  Each linear constraint restricts us to a half-space of $\cF$, and together these half-spaces carve out a convex subset $C\subset\cF$.  We would like to determine whether this subset is non-empty.

In the case at hand, we will take our functional $\a$ to be linear combinations of derivatives with respect to the cross-ratios $z,\bar z$ around the crossing symmetric point $z=\bar z = 1/2$,
\be
\label{eq:functionaldefinition}
\a\p{\begin{array}{c}
f_1\\
f_2\\
f_3
\end{array}
} &=&
\sum_{i=1}^3\sum_{0\leq m+n < 2k} \left.a^i_{m,n}\ptl_z^m\ptl_{\bar z}^n f_i(z,\bar z)\right|_{z=\bar z = 1/2}.
\ee
The parameter $k$ controls the dimension of the space of linear functionals that we search over.  Note that truncating  this search space leads to valid, though possibly suboptimal, bounds.  As we increase $k$, our bounds get better and better, converging to an optimal bound as $k\to \oo$.

\subsection{Formulation as a Semidefinite Program}
\label{sec:formulation}

A key difficulty in our convex optimization problem is that we have an infinite number of constraints on $\a$ --- one for each $\cO$ which could appear in the OPE.  We must impose $\a(F_{R,\De,\ell})\geq 0$ for all representations $R$, dimensions $\De$, and spins $\ell$ obeying our assumptions about the spectrum.  An efficient way to deal with this infinity is to reformulate our problem as a {\it semidefinite program}, which can include constraints of the form:
\be
\label{eq:polynomialinequalities}
\a(P_i(x))\textrm{ for all $x\geq 0$}, \textrm{ where $P_i(x)$ are polynomials in $x$.}
\ee
Systems of these inequalities can be solved efficiently using interior point methods.

To write our constraints $\a(V)\geq 0$ in this form, it suffices to find an approximation
\be
\ptl_z^m\ptl_{\bar z}^n g_{\De,\ell}(z,\bar z)|_{z=\bar z=1/2} &\approx& \chi_\ell(\De)P^{(m,n)}_{\ell}(\De),
\label{eq:polynomialapproximation}
\ee
where $\chi_\ell(\De)$ are positive functions, and $P^{(m,n)}_{\ell}(\De)$ are polynomials.  (Crucially, $\chi_\ell(\De)$ is independent of $m,n$.)  Indeed, assuming Eq.~(\ref{eq:polynomialapproximation}), and combining Eqs.~(\ref{eq:vectordefinitions}) and (\ref{eq:functionaldefinition}), we see that
\be
\a(V_{R,\De,\ell})\geq0 &\textrm{ if and only if }& \sum_{m,n,i} a_{m,n}^i P^{(m,n)}_{R,\ell,i}(\De)\geq 0,
\ee
for polynomials $P^{(m,n)}_{R,\ell,i}(\De)$.  The dimensions $\De$ satisfy bounds $\De\geq \De_{\min,\ell}$, so writing $\De=\De_{\min,\ell}+x$ yields a set of inequalities in the form~(\ref{eq:polynomialinequalities}).

In \cite{Poland:2011ey}, special analytic expressions for conformal blocks in even dimensions \cite{DO1,DO2} were used to derive approximations of the form~(\ref{eq:polynomialapproximation}), which proved sufficient for applying semidefinite programming to even dimensional CFTs.  These approximations worked surprisingly well, but it was unclear how to generalize the techniques to CFTs in odd (or fractional) space-time dimension.

In the next section, we will show that the existence of approximations~(\ref{eq:polynomialapproximation}) in any space-time dimension follows naturally from representation theory of the conformal group.  This is sufficient for formulating our optimization problem as a semidefinite program, which can then be solved using one of the many freely available semidefinite program solvers.  We give details of our implementation using the solver \texttt{SDPA-GMP} in Appendix~\ref{app:semidefinite}.

\section{Rational Representations for Conformal Blocks}
\label{sec:rational}

\subsection{Why Rational Approximations Exist}

\begin{figure}
\centering
\includegraphics{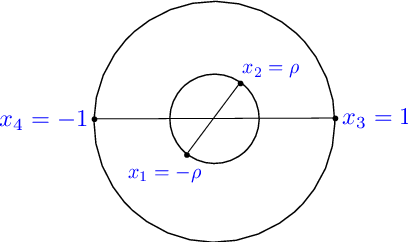}
\caption{Configuration of points for radial quantization in the $\rho$-coordinate \cite{Hogervorst:2013sma}.}
\label{fig:rhocoordinate}
\end{figure}

To compute CFT bounds using semidefinite programming, we need precise, systematic approximations for conformal blocks $g_{\De,\ell}$ in terms of positive functions times polynomials in $\De$, or equivalently positive functions times rational functions of $\De$ with positive denominator.  The existence of such approximations follows from conformal representation theory.  Recall that the conformal block $g_{\De,\ell}$ is a sum over states in radial quantization
\be
\frac{g_{\De,\ell}(u,v)}{x_{12}^{2\De_\f}x_{34}^{2\De_\f}} &=& \sum_{\a=\cO,\,P\cO,\,PP\cO,\,\dots}\frac{\<0|\f(x_1)\f(x_2)|\a\>\<\a|\f(x_3)\f(x_4)|0\>}{\<\a|\a\>} ,
\label{eq:blockassumoverstates}
\ee
where $\cO$ is a conformal primary of dimension $\De$ and spin $\ell$, and $\a$ runs over $\cO$ and all of its descendants.  It will be convenient for our discussion to use the radial coordinates of \cite{Hogervorst:2013sma}, where the points $x_i$ are arranged as in Fig.~\ref{fig:rhocoordinate}, and the coordinate $\rho$ is given in terms of cross-ratios by
\bea
\rho = \frac{z}{(1+\sqrt{1-z})^2},\qquad \bar\rho = \frac{\bar z}{(1+\sqrt{1-\bar z})^2}.
\eea

The states $P^{\mu_1}\cdots P^{\mu_n}|\cO\>$ have eigenvalue $\De+n$ under dilatation and can be decomposed into traceless symmetric tensor representations of the rotation group.  It follows that the sum over states~(\ref{eq:blockassumoverstates}) can be written
\bea
\label{eq:gegenbauerexpansion}
g_{\De,\ell}(r,\eta) &=& \sum_{n=0}^\oo \sum_j B_{n,j} r^{\De+n} \frac{j!}{(2\nu)_j}C_j^\nu(\eta) ,
\eea
where $r = |\rho|$, $\eta = \cos\th = (\rho + \bar\rho)/2|\rho|$, $\nu = D/2-1$, and the $C_j^\nu(\eta)$ are Gegenbauer polynomials.  The coefficients $B_{n,j}$ can be computed straightforwardly by solving the conformal Casimir equation
\be
\label{eq:casimireqn}
\cD g_{\De,\ell} = C_{\De,\ell}g_{\De,\ell},\qquad C_{\De,\ell}=\De(\De-D)+\ell(\ell+D-2),
\ee
term by term around $r=0$.  Here, $\cD$ is a second-order differential operator in the cross-ratios $u,v$, representing the action of the quadratic Casimir of the conformal group on a four-point function,
\be
\cD g(u,v) &=& \frac 1 2 (L^{(1)}_{AB}+L^{(2)}_{AB})(L^{(1)AB}+L^{(2)AB})g(u,v).
\ee
Further details can be found in \cite{Hogervorst:2013sma}.

Solving the Casimir equation to second order in $r$, we find \cite{Hogervorst:2013sma}
\begin{align}
B_{2,\ell-2}&=\frac{\ell(\ell-1)(\De-\ell-2\nu)}{2(\ell+\nu-1)(\ell+\nu)(\De-\ell-2\nu+1)}, & B_{2,\ell}=\nu\frac{\De\nu(\nu-1)+(\De-1)\ell(\ell+2\nu)}{(\De-\nu)(\ell+\nu+1)(\ell+\nu-1)},\nn\\
B_{2\ell+2}&=\frac{(\De+\ell)(\ell+2\nu)(\ell+2\nu+1)}{2(\De+\ell+1)(\ell+\nu)(\ell+\nu+1)}.
\end{align}

Let us make some comments about these coefficients.  First, $B_{n,j}$ is a rational function of $\De$.  This is a simple consequence of the Casimir equation~(\ref{eq:casimireqn}), but it follows more directly from the expression for $g_{\De,\ell}$ as a sum over states.  Each term in the numerator and denominator of~(\ref{eq:blockassumoverstates}) can be computed from the action of the conformal algebra on $|\cO\>$.  This action is polynomial in $\De$, so $B_{n,j}$ is rational in $\De$.

Secondly, the denominators of $B_{n,j}$ are positive, as long as $\De$ obeys the unitarity bound~(\ref{eq:unitaritybound}).  This is because poles in $B_{n,j}$ occur when a state $|\a\>$ becomes null.  However, the absence of null or negative-norm states in a conformal multiplet is precisely what defines the unitarity bound.

These facts are exactly what we need.  We can obtain good approximations to conformal blocks by truncating the expansion~(\ref{eq:gegenbauerexpansion}) at some high order $r^{\De+N}$.  Each term in the resulting expression will be a rational function of $\De$, times an overall factor of $r^{\De}$.  The error is na\"ively of order $r^N$, which is quite small at the crossing-symmetric point $r=3-2\sqrt 2\approx 0.17$.  Indeed, this na\"ive estimate is correct, since it can be shown that the coefficients $B_{n,j}$ are uniformly bounded as a function of $\De$ for all $n,j$.

The cost of keeping more terms in the series expansion~(\ref{eq:gegenbauerexpansion}) is that the degree of the resulting rational approximation grows.  As we'll see in the next subsection, this growth is slow and under control.  Further, additional tricks can improve the rational approximation significantly without increasing the degree.  We discuss these in Appendix~\ref{app:truncatingrationalapproximations}.

\subsection{Poles in $\De$ and Recursion Relations for $g_{\De,\ell}$}

We can be more precise about the structure of our rational approximations by exploiting an idea of Zamolodchikov, originally applied to Virasoro blocks in 2 dimensions~\cite{Zamolodchikov:1985ie,Zamolodchikov:1987}.  Recall that poles in $g_{\De,\ell}$ as a function of $\De$ occur precisely at special values $\De=\De_*$ where some state $|\a\>=P^n |\cO\>$ in the conformal multiplet of $|\cO\>$ becomes null.  When this happens, all the descendants of $|\a\>$ become null as well, and together they form a nontrivial sub-representation of the (now reducible) multiplet of $|\cO\>$.  Since the pole in $\De$ gets contributions from all states in this sub-representation, its residue is proportional to a conformal block for a primary with the same dimension and spin as $|\a\>$:
\be
g_{\De,\ell} &\sim& \frac{c_\a}{\De-\De_*}g_{\De_{\a}, \ell_{\a}}\textrm{\quad as \quad}\De\to\De_*,
\ee
where $\De_{\a}=\De_*+n$, and $c_\a$ is a coefficient which is independent of conformal cross-ratios.

Poles in $\De$ determine $g_{\De,\ell}$ up to a function which is analytic on the entire complex plane.  Thus, we can write
\be
\label{eq:recursiong}
g_{\De,\ell}(r,\eta) &=& g^{(\oo)}_{\ell}(\De,r,\eta) + \sum_i \frac{c_i}{\De-\De_i}g_{\De_i+n_i,\ell_i}(r,\eta),
\ee
where $g^{(\oo)}_{\ell}(\De,r,\eta)$ is an entire function of $\De$.  The block $g_{\De,\ell}$ has an essential singularity of the form $r^\De$ as $\De\to\oo$.  Stripping this off, we have
\be
h_{\De,\ell}(r,\eta) &\equiv& r^{-\De}g_{\De,\ell}(r,\eta),\\
h_{\De,\ell}(r,\eta) &=& h^{(\oo)}_\ell(r,\eta) + \sum_i \frac{c_i r^{n_i}}{\De-\De_i}h_{\De_i+n_i,\ell_i}(r,\eta).
\label{eq:recursionh}
\ee
Note that the entire function $h_\ell^{(\oo)}(r,\eta)=\lim_{\De\to\oo} h_{\De,\ell}(r,\eta)$ is now independent of $\De$, since it has no singularities as $\De\to\oo$.

With the recursion relation~(\ref{eq:recursionh}), we can be more explicit about the form of our positive-times-polynomial approximation~(\ref{eq:polynomialapproximation}).  Truncating~(\ref{eq:recursionh}) to a finite number of poles $\De_i$, it's clear that
\be
\ptl_z^m\ptl_{\bar z}^n g_{\De,\ell}(z,\bar z)|_{z=\bar z=1/2} \approx \chi_\ell(\De)P^{(m,n)}_{\ell}(\De),\qquad\textrm{ where }\quad
\chi_\ell(\De) \equiv \frac{r^{\De}}{\prod_i(\De-\De_i)},
\ee
and $r=3-2\sqrt 2$ is evaluated at the crossing-symmetric point.

Let us now determine the data entering~(\ref{eq:recursionh}).  The function $h_{\ell}^{(\oo)}$ can be computed easily by solving the conformal Casimir equation to leading order in $\De$,
\be
h_{\ell}^{(\oo)}(r,\eta) &=& \frac{\ell!}{(2\nu)_\ell}\frac{C_\ell^\nu(\eta)}{(1-r^2)^\nu\sqrt{(1+r^2)^2-4r^2\eta^2}}.
\ee
In principle, the pole positions $\De_i$ and coefficients $c_i$ are determined by the conformal algebra.  It would be interesting to compute them directly.  In practice, by inspecting the solution to the conformal Casimir equation, we find that there are three series of poles, described in Table~\ref{tab:polepositions}, whose coefficients are as follows:
\begin{table}
\centering
\begin{tabular}{c|c|c|cl}
$n_i$ & $\De_i$ & $\ell_i$ & $c_i$\\
\cline{1-4}
$2k$ & $1-\ell-2k$ & $\ell+2k$ & $c_1(k)$  & \quad $k=1,2,\dots$\\
$2k$ & $1+\nu-k$   & $\ell$    & $c_2(k)$  & \quad $k=1,2,\dots$\\ 
$2k$ & $1+\ell+2\nu-2k$ & $\ell-2k$ & $c_3(k)$ & \quad $k=1,2,\dots,\lfloor \ell/2 \rfloor$
\end{tabular}
\caption{The positions of poles of $g_{\De,\ell}$ in $\Delta$ and their associated data.  There are three types of poles, corresponding to the three rows in the table.  The first two types exist for all positive integer $k$, while the third type exists for positive integer $k\leq \lfloor \ell/2\rfloor$.  The coefficients $c_1(k), c_2(k), c_3(k)$ are given in Eqs.~(\ref{eq:poleresidues}).}
\label{tab:polepositions}
\end{table}

\be
\label{eq:poleresidues}
c_1(k) &=& -\frac{k(2k)!^2}{2^{4k-1}k!^4}\frac{(\ell+2\nu)_{2k}}{(\ell+\nu)_{2k}},\nn\\
c_2(k) &=& -\frac{k(\nu+\ell-k)(\nu)_k(1-\nu)_k\p{\frac{\nu+\ell+1-k}{2}}_k^2}{k!^2(\nu+\ell+k)\p{\frac{\nu+\ell-k}{2}}_k^2},\nn\\
c_3(k) &=& -\frac{k(2k)!^2}{2^{4k-1}k!^4}\frac{(1+\ell-2k)_{2k}}{(1+\nu+\ell-2k)_{2k}}.
\ee

The recursion relation Eq.~(\ref{eq:recursionh}), with poles listed in Table~\ref{tab:polepositions}, reveals another fact that proves useful for our implementation: when we truncate the series expansion for $h_{\De,\ell}$ at order $r^{N}$, the degree of the resulting rational approximation grows like $N$.  Not only is each coefficient $B_{n,j}$ a rational function, but the number of new factors which enter the denominator as we increase $N\to N+2$ is either 2 or 3 (depending on whether $N\leq \ell$).  In practice, this means that we can compute the expansion~(\ref{eq:gegenbauerexpansion}) to extremely high order without incurring too much of a performance hit from dealing with large degree polynomials.

Once we know the residues $c_i(k)$, our recursion relation~(\ref{eq:recursionh}) provides an extremely efficient way to compute conformal blocks, either analytically in a series expansion, or numerically (for any number of derivatives around any point $(r,\eta)$).  It would be very interesting to generalize these ideas to other four-point functions --- for example, external scalars $\f_i$ with different dimensions $\De_i$, or external operators with spin.  Although much progress has been made computing conformal blocks for these types of operators using series manipulations \cite{DO1,DO2,DO3,ElShowk:2012ht,Hogervorst:2013sma}, derivative relations \cite{Costa:2011mg,Costa:2011dw}, and conformal integrals \cite{SimmonsDuffin:2012uy}, we believe the residues $c_i$ are the most convenient and directly useful data for numerical bootstrap applications.

\section{Results and Comparison to $O(N)$ Vector Models}
\label{sec:results}

In this section we will present our results from the bootstrap. We focus our attention on computing bounds on the lowest singlet dimension $\Delta_S$, the lowest symmetric tensor dimension $\Delta_T$, and the central charge $c$, as a function of the external scalar dimension $\Delta_\phi$.

Let us begin by summarizing what is known about these quantities in the $O(N)$ vector models. In these theories, results are typically phrased in terms of critical exponents, rather than scaling dimensions. Concretely, $\Delta_{\phi}$ is related to the critical exponent $\eta$ via $\Delta_\phi = 1/2 + \eta/2$.  The dimension of the $O(N)$ singlet operator $S$ is related to the critical exponent $\nu$ via $\Delta_S = 3-1/\nu$. Finally, the dimension of the $O(N)$ symmetric tensor operator $T$ is related to the crossover exponents $\phi_c$ and $\eta_c$, through the relations $\Delta_T = 3-\phi_c/\nu = 1+\eta_c$. In Table~\ref{table:prev} we show the most accurate determinations of these dimensions that we have found in the literature for $N=1,2,3,4,5,6$ (with $N=1$ being the 3D Ising model). 

\begin{table}[]
\centering
\begin{tabular}{c c c c} \hline\hline
 \hspace{0.4cm} $N$ \hspace{0.4cm}  & \hspace{0.4cm} $\De_{\phi} =1/2+\eta/2$ \hspace{0.4cm}  & \hspace{0.4cm}  $\Delta_S=3-1/\nu$ \hspace{0.4cm}  & \hspace{0.4cm}  $\Delta_T=3-\phi_c/\nu=1+\eta_c$ \hspace{0.4cm}  \\
\cline{1-4}
$1$ & $0.51813(5)$   \cite{Hasenbusch:2011yya} & $1.41275(25)$ \cite{Hasenbusch:2011yya} & --  \\
        & $0.51819(7)$ \cite{Campostrini:2002cf} & $1.4130(4)$ \cite{Campostrini:2002cf} & -- \\
$2$ & $0.51905(10)$ \cite{Campostrini:2006ms} & $1.51124(22)$ \cite{Campostrini:2006ms} & 1.237(4) \cite{Calabrese:2004ca}  \\ 
$3$ & $0.51875(25)$ \cite{Campostrini:2002ky} & $1.5939(10)$ \cite{Campostrini:2002ky} & 1.211(3) \cite{Calabrese:2004ca} \\
$4$ & $0.51825(50)$ \cite{Hasenbusch:2000ph} & $1.6649(35)$ \cite{Hasenbusch:2000ph} & 1.189(2) \cite{Calabrese:2004ca} \\
$5$ & $0.5155(15)$ \cite{Butti:2004qr} & $1.691(7)$ \cite{Butti:2004qr} & 1.170(2) \cite{Calabrese:2004ca} \\
$6$ & $0.5145(15)$ \cite{Butti:2004qr} & $1.733(8)$ \cite{Butti:2004qr}& -- \\
\hline\hline
\end{tabular}
\caption{Previous determinations of operator dimensions in the $O(N)$ vector models.}
\label{table:prev}
\end{table}

These quantities have also been computed in the large $N$ limit. $\Delta_{\phi}$ is known to order $1/N^3$ while $\Delta_S$ has only been computed to order $1/N^2$ (see~\cite{Moshe:2003xn} and references therein). The crossover exponent connected to $\Delta_T$ was also computed to order $1/N^2$ in~\cite{Gracey:2002qa}. The results are:
\bea
\label{eq:largeNdimensions}
\Delta_{\phi} &=& \frac{1}{2} + \frac{4}{3\pi^2} \frac{1}{N} -\frac{256}{27\pi^4} \frac{1}{N^2} \nonumber\\
&&+ \frac{32\left(-3188+3\pi^2(-61+108\log(2))-3402 \zeta(3) \right)}{243\pi^6}  \frac{1}{N^3} 
 + \cO \left(\frac{1}{N^4}\right) \nonumber\\
\Delta_{S} &=& 2 - \frac{32}{3\pi^2} \frac{1}{N} + \frac{32(16-27\pi^2)}{27\pi^4} \frac{1}{N^2} + \cO \left(\frac{1}{N^3}\right) \nonumber\\
\Delta_{T} &=& 1 + \frac{32}{3\pi^2} \frac{1}{N} - \frac{512}{27\pi^4} \frac{1}{N^2} + \cO \left(\frac{1}{N^3}\right) .
\eea
Finally, let us mention that the leading correction to the central charge $c$ was computed in~\cite{Petkou:1994ad,Petkou:1995vu} to be
\bea
\label{eq:largeNcentralcharge}
\frac{c}{N c_{\textrm{free}}} = 1 - \frac{40}{9\pi^2} \frac{1}{N} + \cO \left(\frac{1}{N^2}\right) ,
\eea
where $c_{\textrm{free}} = D/(D-1)$ is the central charge of a free scalar field.

\subsection{Bounds on $O(N)$ Singlets}
\label{sec:boundsonsinglets}
Now we determine a general bound on the $O(N)$ singlet operator dimension $\Delta_S$ following the procedure described in section~\ref{sec:convex}. We assume there is a gap in the CFT spectrum so that all singlet scalar operators have dimension greater than $\Delta_S$, all symmetric tensor scalars have dimension greater than 1, and the dimensions of all the other operators are constrained only by the unitarity conditions. Note that due to the assumption on symmetric tensor scalars this is not the most general bound. However, we found that this mild assumption improves numerical stability while not significantly affecting the bound on $\Delta_S$ -- moreover the assumption is certainly satisfied for $O(N)$ vector models, as can be seen from previous determinations of the operator dimensions (see Table~\ref{table:prev}). 

\begin{figure}[]
\begin{center}
\begin{psfrags}
\def\PFGstripminus-#1{#1}%
\def\PFGshift(#1,#2)#3{\raisebox{#2}[\height][\depth]{\hbox{%
  \ifdim#1<0pt\kern#1 #3\kern\PFGstripminus#1\else\kern#1 #3\kern-#1\fi}}}%
\providecommand{\PFGstyle}{}%
%
\psfrag{deltaPhi}[cl][cl]{\PFGstyle $\De_\f$}%
\psfrag{deltaS}[bc][bc]{\PFGstyle $\De_{S}$}%
\psfrag{Ising}[cc][cc]{\PFGstyle $\quad\text{Ising}$}%
\psfrag{O10}[cc][cc]{\PFGstyle $\quad\ O(10)$}%
\psfrag{O20}[cc][cc]{\PFGstyle $\quad\ O(20)$}%
\psfrag{O2}[cc][cc]{\PFGstyle $\quad O(2)$}%
\psfrag{O3}[cc][cc]{\PFGstyle $\quad O(3)$}%
\psfrag{O4}[cc][cc]{\PFGstyle $\quad O(4)$}%
\psfrag{O5}[cc][cc]{\PFGstyle $\quad O(5)$}%
\psfrag{O6}[cc][cc]{\PFGstyle $\quad O(6)$}%
\psfrag{ONSingletB}[bc][bc]{\PFGstyle $\text{$O(N)$ Singlet Bounds}$}%
\psfrag{x0}[tc][tc]{\PFGstyle $0$}%
\psfrag{x11}[tc][tc]{\PFGstyle $1$}%
\psfrag{x25}[tc][tc]{\PFGstyle $0.25$}%
\psfrag{x505}[tc][tc]{\PFGstyle $0.505$}%
\psfrag{x515}[tc][tc]{\PFGstyle $0.515$}%
\psfrag{x51}[tc][tc]{\PFGstyle $0.51$}%
\psfrag{x525}[tc][tc]{\PFGstyle $0.525$}%
\psfrag{x52}[tc][tc]{\PFGstyle $0.52$}%
\psfrag{x535}[tc][tc]{\PFGstyle $0.535$}%
\psfrag{x53}[tc][tc]{\PFGstyle $0.53$}%
\psfrag{x5}[tc][tc]{\PFGstyle $0.5$}%
\psfrag{x75}[tc][tc]{\PFGstyle $0.75$}%
\psfrag{xm11}[tc][tc]{\PFGstyle $-1$}%
\psfrag{xm25}[tc][tc]{\PFGstyle $-0.25$}%
\psfrag{xm5}[tc][tc]{\PFGstyle $-0.5$}%
\psfrag{xm75}[tc][tc]{\PFGstyle $-0.75$}%
\psfrag{y0}[cr][cr]{\PFGstyle $0$}%
\psfrag{y11}[cr][cr]{\PFGstyle $1$}%
\psfrag{y121}[cr][cr]{\PFGstyle $1.2$}%
\psfrag{y141}[cr][cr]{\PFGstyle $1.4$}%
\psfrag{y161}[cr][cr]{\PFGstyle $1.6$}%
\psfrag{y181}[cr][cr]{\PFGstyle $1.8$}%
\psfrag{y21}[cr][cr]{\PFGstyle $2$}%
\psfrag{y221}[cr][cr]{\PFGstyle $2.2$}%
\psfrag{y25}[cr][cr]{\PFGstyle $0.25$}%
\psfrag{y5}[cr][cr]{\PFGstyle $0.5$}%
\psfrag{y75}[cr][cr]{\PFGstyle $0.75$}%
\psfrag{ym11}[cr][cr]{\PFGstyle $-1$}%
\psfrag{ym25}[cr][cr]{\PFGstyle $-0.25$}%
\psfrag{ym5}[cr][cr]{\PFGstyle $-0.5$}%
\psfrag{ym75}[cr][cr]{\PFGstyle $-0.75$}%
\includegraphics[width=0.9\textwidth]{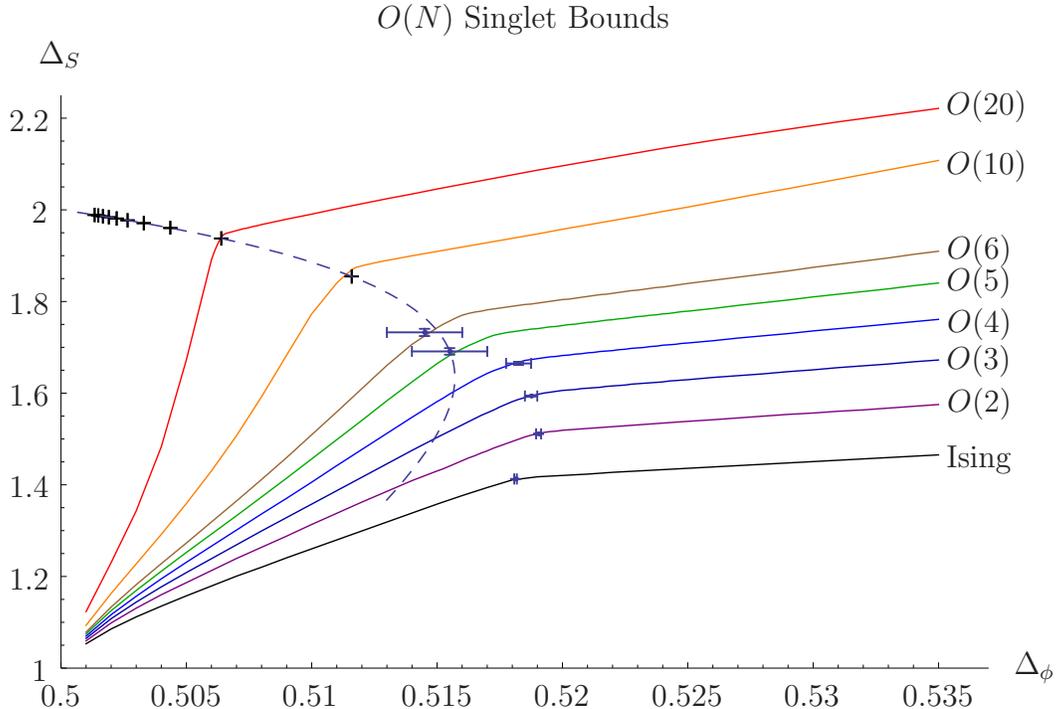}
\end{psfrags}
\caption{Upper bounds on the dimension of the lowest dimension singlet $S$ in the $\f\x\f$ OPE, where $\f$ transforms as a vector under an $O(N)$ global symmetry group.  Here, we show $N=1,2,3,4,5,6,10,20$.  The blue error bars represent the best available analytical and Monte Carlo determinations of the operator dimensions $(\De_\f,\De_{S})$ in the $O(N)$ vector models for $N=1,2,3,4,5,6$ (with $N=1$ being the 3D Ising Model).  The black crosses show the predictions in Eq.~(\ref{eq:largeNdimensions}) from the large-$N$ expansion for $N=10,20,..., 100$.  In this expansion, $\De_\f$ has been determined to three-loop order, while $\De_{S}$ is at two-loop order.  The dashed line interpolates the large-$N$ prediction for $N\in(4,\oo)$.}
\label{fig:singletbounds}
\end{center}
\end{figure}

The boundaries for the allowed values of $\Delta_S$ as a function of $\Delta_\phi$ are shown in Fig.~\ref{fig:singletbounds}.  These bounds are determined by a bisection search in $\De_S$ to within $10^{-3}$. The parameter $k$ of section~\ref{sec:formulation}, controlling the number of derivatives in the functional $\alpha$, is set to $k=10$ everywhere.  For a given $N$, only the values of $\Delta_S$ below the corresponding solid line are allowed.

In Fig.~\ref{fig:singletbounds} we see that the bounds on $\Delta_S$ grow monotonically from $\Delta_S=1$ at $\Delta_\phi=0.5$, the point corresponding to the non-interacting theory. At a certain value of $\Delta_\phi$, each boundary line exhibits a change in the slope. This type of behavior was already discussed for the Ising model bound in Ref.~\cite{ElShowk:2012ht}, where it was found that the change in slope occurs at the values of $\Delta_\phi$ and $\Delta_S$ corresponding to Ising model point. The changes in slopes are somewhat less sharp for the $O(N)$ theories; however, they occur in the vicinity of the O(N) vector model points, shown as points with error bars in Fig.~\ref{fig:singletbounds}. These points always seem to lie very close to the boundary; in fact it is possible to rigorously reduce the previous error bars using the bound we obtained.

For large values of $N$, we can also compare the bound with the results for $\Delta_\phi$ and $\Delta_S$ obtained using the $1/N$ expansion, Eq.~\ref{eq:largeNdimensions}. The $1/N$ expansion results (shown as black crosses in Fig.~\ref{fig:singletbounds}) are consistent with our bound at large $N$. Moreover, the change in the slope of the boundary line is sharper for large $N$ and it occurs very close to the $O(N)$ vector model points. 

\subsection{Bounds on $O(N)$ Symmetric Tensors}
\begin{figure}[h!]
\begin{center}
\begin{psfrags}
\def\PFGstripminus-#1{#1}%
\def\PFGshift(#1,#2)#3{\raisebox{#2}[\height][\depth]{\hbox{%
  \ifdim#1<0pt\kern#1 #3\kern\PFGstripminus#1\else\kern#1 #3\kern-#1\fi}}}%
\providecommand{\PFGstyle}{}%
%
\psfrag{deltaPhi}[cl][cl]{\PFGstyle $\De_\f$}%
\psfrag{deltaT}[bc][bc]{\PFGstyle $\De_{T}$}%
\psfrag{O10}[cc][cc]{\PFGstyle $\quad\ O(10)$}%
\psfrag{O20}[cc][cc]{\PFGstyle $\quad\ O(20)$}%
\psfrag{O2}[cc][cc]{\PFGstyle $\quad O(2)$}%
\psfrag{O3}[cc][cc]{\PFGstyle $\quad O(3)$}%
\psfrag{O4}[cc][cc]{\PFGstyle $\quad O(4)$}%
\psfrag{O5}[cc][cc]{\PFGstyle $\quad O(5)$}%
\psfrag{O6}[cc][cc]{\PFGstyle $\quad O(6)$}%
\psfrag{ONSymmetri}[bc][bc]{\PFGstyle $\text{$O(N)$ Symmetric Tensor Bounds}$}%
\psfrag{x0}[tc][tc]{\PFGstyle $0$}%
\psfrag{x11}[tc][tc]{\PFGstyle $1$}%
\psfrag{x25}[tc][tc]{\PFGstyle $0.25$}%
\psfrag{x505}[tc][tc]{\PFGstyle $0.505$}%
\psfrag{x515}[tc][tc]{\PFGstyle $0.515$}%
\psfrag{x51}[tc][tc]{\PFGstyle $0.51$}%
\psfrag{x525}[tc][tc]{\PFGstyle $0.525$}%
\psfrag{x52}[tc][tc]{\PFGstyle $0.52$}%
\psfrag{x535}[tc][tc]{\PFGstyle $0.535$}%
\psfrag{x53}[tc][tc]{\PFGstyle $0.53$}%
\psfrag{x5}[tc][tc]{\PFGstyle $0.5$}%
\psfrag{x75}[tc][tc]{\PFGstyle $0.75$}%
\psfrag{xm11}[tc][tc]{\PFGstyle $-1$}%
\psfrag{xm25}[tc][tc]{\PFGstyle $-0.25$}%
\psfrag{xm5}[tc][tc]{\PFGstyle $-0.5$}%
\psfrag{xm75}[tc][tc]{\PFGstyle $-0.75$}%
\psfrag{y0}[cr][cr]{\PFGstyle $0$}%
\psfrag{y1051}[cr][cr]{\PFGstyle $1.05$}%
\psfrag{y111}[cr][cr]{\PFGstyle $1.1$}%
\psfrag{y1151}[cr][cr]{\PFGstyle $1.15$}%
\psfrag{y11}[cr][cr]{\PFGstyle $1$}%
\psfrag{y121}[cr][cr]{\PFGstyle $1.2$}%
\psfrag{y1251}[cr][cr]{\PFGstyle $1.25$}%
\psfrag{y131}[cr][cr]{\PFGstyle $1.3$}%
\psfrag{y25}[cr][cr]{\PFGstyle $0.25$}%
\psfrag{y5}[cr][cr]{\PFGstyle $0.5$}%
\psfrag{y75}[cr][cr]{\PFGstyle $0.75$}%
\psfrag{ym11}[cr][cr]{\PFGstyle $-1$}%
\psfrag{ym25}[cr][cr]{\PFGstyle $-0.25$}%
\psfrag{ym5}[cr][cr]{\PFGstyle $-0.5$}%
\psfrag{ym75}[cr][cr]{\PFGstyle $-0.75$}%
\includegraphics[width=0.9\textwidth]{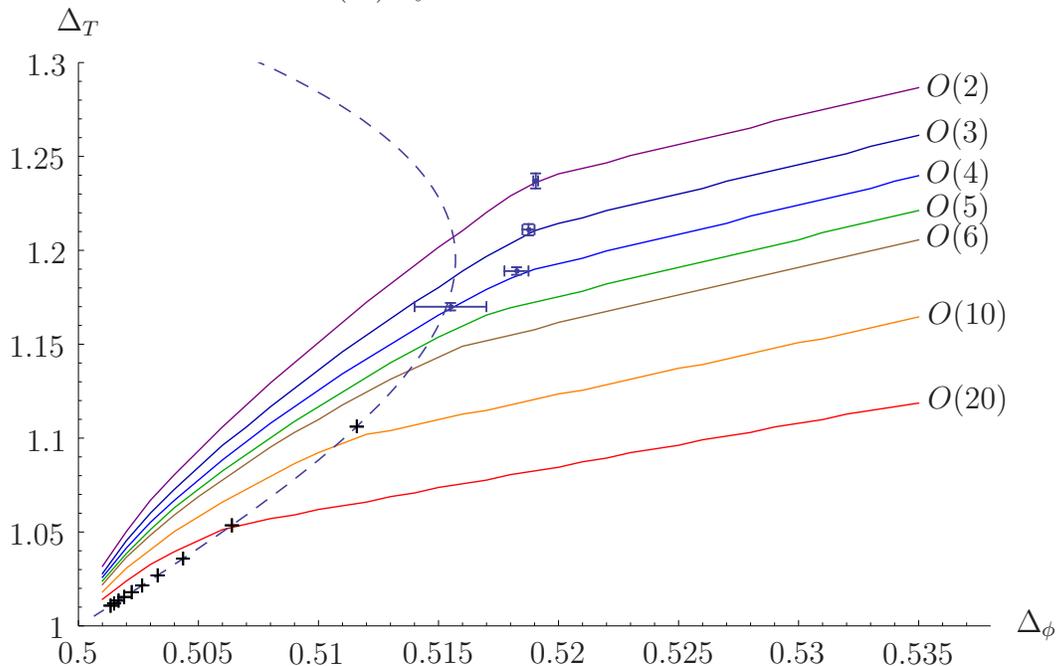}
\end{psfrags}
\caption{Upper bounds on the dimension of the lowest dimension symmetric tensor $T$ in the $\f\x\f$ OPE, where $\f$ transforms as a vector under an $O(N)$ global symmetry group, for $N=2,3,4,5,6,10,20$.  We additionally assume that the lowest dimension singlet $S$ has $\De_S\geq 1$.  The blue error bars represent the best available analytical and Monte Carlo determinations of the operator dimensions $(\De_\f,\De_{T})$ in the $O(N)$ vector models for $N=2,3,4,5$.  Note in particular that previous predictions for the $O(5)$ model are essentially ruled out by our bounds.  The black crosses show the predictions in Eq.~(\ref{eq:largeNdimensions}) from the large-$N$ expansion for $N=10,20,..., 100$.  The dashed line interpolates the large-$N$ prediction for $N\in(4,\oo)$.}

\label{fig:tensorbounds}
\end{center}
\end{figure}

Bounds on the dimension of the first symmetric tensor scalar operator $\Delta_T$ are obtained following a similar procedure as for the singlet bounds. We assume that all the symmetric tensor scalars have dimension greater than $\Delta_T$, all singlet scalars have dimension greater than 1, and all other operator dimensions satisfy unitarity conditions. The allowed values of $\Delta_T$ as a function of $\Delta_\phi$ are shown in Fig.~\ref{fig:tensorbounds} --- only the values of $\Delta_T$ below the solid line are allowed. Just like in the case of the singlet operators, the boundary lines start at the free field point, $\Delta_\phi=0.5$, $\Delta_T=1$ and grow monotonically. The change of the slope is more gradual than for the singlet bound. The $O(N)$ vector model points from Table~\ref{table:prev} appear to be consistent with the bound to within error bars for $N=2,3,4$, with some mild tension between the $N=5$ bound and the quoted results from~\cite{Butti:2004qr} and~\cite{Calabrese:2004ca}. 

At large $N$ the symmetric tensor dimension $\Delta_T$ approaches $1$ in the $O(N)$ model. This is reflected in the fact that our bounds become more and more constraining as $N \rightarrow \infty$. Moreover, we find that the order $1/N^2$ computation of $\Delta_T$ in Eq.~\ref{eq:largeNdimensions} becomes very close to our bound (and occurs near changes of slope) at large but finite values of $N$, shown in black crosses at $N=10,20,...$ in Fig.~\ref{fig:tensorbounds}.

\subsection{Bounds on the Central Charge}
The central charge $c$ is defined as a coefficient in the two-point correlation function of the canonically normalized stress tensor:
\be
\label{eq:stresstensor}
\langle T_{\mu \nu}(x_1) T_{\rho\sigma}(x_2) \rangle = \frac{c}{S_D^2} \frac{1}{x_{12}^{2D}} \left\{ \frac{1}{2} \left[ I_{\mu\rho}(x_{12}) I_{\nu\sigma} (x_{12}) +I_{\mu\sigma}(x_{12}) I_{\nu\rho} (x_{12})  \right] - \frac{1}{D}\eta_{\mu\nu}\eta_{\rho\sigma} \right\},
\ee
where $S_D = 2\pi^{D/2}/\Gamma(D/2) $. In our notation, the central charge is related to the OPE coefficient of the stress tensor $\lambda_{S,3,2}$ by
\be
\label{eq:centralchargecoefficient}
\lambda_{S,3,2}^2 = \frac{\Delta_\phi^2}{c/c_{\text{free}}},
\ee
where $c_{\text{free}} = D/(D-1)$ is the central charge of a free scalar field. We can find an upper bound on this OPE coefficient as follows. Rewrite the sum rule (\ref{eq:unitopisolated}), separating the contribution of the stress tensor:
\be
\label{eq:stresstensorisolated}
\lambda_{S,3,2}^2 V_{S,3,2} = - V_{\text{unit}} - \sum_{\cO \neq T_{\mu\nu}}\lambda_{\cO}^2 V_\cO .
\ee
Applying a functional $\alpha$ such that  $\alpha (V_\cO) \ge 0$ for all operators in the spectrum other than unit operator and normalized so that $\alpha(V_{S,3,2}) = 1$, Eq.~(\ref{eq:stresstensorisolated}) then yields the inequality
\be
\lambda_{S,3,2}^2  \le -\alpha ( V_{\text{unit}} ) .
\ee
Finding the functional $\alpha$ that minimizes $-\alpha(V_{\text{unit}})$ then gives the strongest upper bound on the OPE coefficient. By Eq.~(\ref{eq:centralchargecoefficient}) this implies a lower bound on the central charge.

\begin{figure}[h!]
\begin{center}
\begin{psfrags}
\def\PFGstripminus-#1{#1}%
\def\PFGshift(#1,#2)#3{\raisebox{#2}[\height][\depth]{\hbox{%
  \ifdim#1<0pt\kern#1 #3\kern\PFGstripminus#1\else\kern#1 #3\kern-#1\fi}}}%
\providecommand{\PFGstyle}{\small}%
%
\psfrag{deltaPhi}[cl][cl]{\PFGstyle $\De_\f$}%
\psfrag{cOverCFree}[bc][bc]{\PFGstyle $c/Nc_\mathrm{free}$}%
\psfrag{Ising}[cc][cc]{\PFGstyle $\quad\text{Ising}$}%
\psfrag{O10}[cc][cc]{\PFGstyle $\quad\ O(10)$}%
\psfrag{O20}[cc][cc]{\PFGstyle $\quad\ O(20)$}%
\psfrag{O2}[cc][cc]{\PFGstyle $\quad O(2)$}%
\psfrag{O3}[cc][cc]{\PFGstyle $\quad O(3)$}%
\psfrag{O4}[cc][cc]{\PFGstyle $\quad O(4)$}%
\psfrag{O5}[cc][cc]{\PFGstyle $\quad O(5)$}%
\psfrag{O6}[cc][cc]{\PFGstyle $\quad O(6)$}%
\psfrag{ONCentralC}[bc][bc]{\PFGstyle $O(N) ~\text{Central Charge Bounds}$}%
\psfrag{x0}[tc][tc]{\PFGstyle $0$}%
\psfrag{x11}[tc][tc]{\PFGstyle $1$}%
\psfrag{x25}[tc][tc]{\PFGstyle $0.25$}%
\psfrag{x505}[tc][tc]{\PFGstyle $0.505$}%
\psfrag{x515}[tc][tc]{\PFGstyle $0.515$}%
\psfrag{x51}[tc][tc]{\PFGstyle $0.51$}%
\psfrag{x525}[tc][tc]{\PFGstyle $0.525$}%
\psfrag{x52}[tc][tc]{\PFGstyle $0.52$}%
\psfrag{x535}[tc][tc]{\PFGstyle $0.535$}%
\psfrag{x53}[tc][tc]{\PFGstyle $0.53$}%
\psfrag{x5}[tc][tc]{\PFGstyle $0.5$}%
\psfrag{x75}[tc][tc]{\PFGstyle $0.75$}%
\psfrag{xm11}[tc][tc]{\PFGstyle $-1$}%
\psfrag{xm25}[tc][tc]{\PFGstyle $-0.25$}%
\psfrag{xm5}[tc][tc]{\PFGstyle $-0.5$}%
\psfrag{xm75}[tc][tc]{\PFGstyle $-0.75$}%
\psfrag{y0}[cr][cr]{\PFGstyle $0$}%
\psfrag{y1051}[cr][cr]{\PFGstyle $1.05$}%
\psfrag{y10251}[cr][cr]{\PFGstyle $1.025$}%
\psfrag{y11}[cr][cr]{\PFGstyle $1$}%
\psfrag{y25}[cr][cr]{\PFGstyle $0.25$}%
\psfrag{y5}[cr][cr]{\PFGstyle $0.5$}%
\psfrag{y75}[cr][cr]{\PFGstyle $0.75$}%
\psfrag{y85}[cr][cr]{\PFGstyle $0.85$}%
\psfrag{y875}[cr][cr]{\PFGstyle $0.875$}%
\psfrag{y925}[cr][cr]{\PFGstyle $0.925$}%
\psfrag{y95}[cr][cr]{\PFGstyle $0.95$}%
\psfrag{y975}[cr][cr]{\PFGstyle $0.975$}%
\psfrag{y9}[cr][cr]{\PFGstyle $0.9$}%
\psfrag{ym11}[cr][cr]{\PFGstyle $-1$}%
\psfrag{ym25}[cr][cr]{\PFGstyle $-0.25$}%
\psfrag{ym5}[cr][cr]{\PFGstyle $-0.5$}%
\psfrag{ym75}[cr][cr]{\PFGstyle $-0.75$}%
\includegraphics[width=0.9\textwidth]{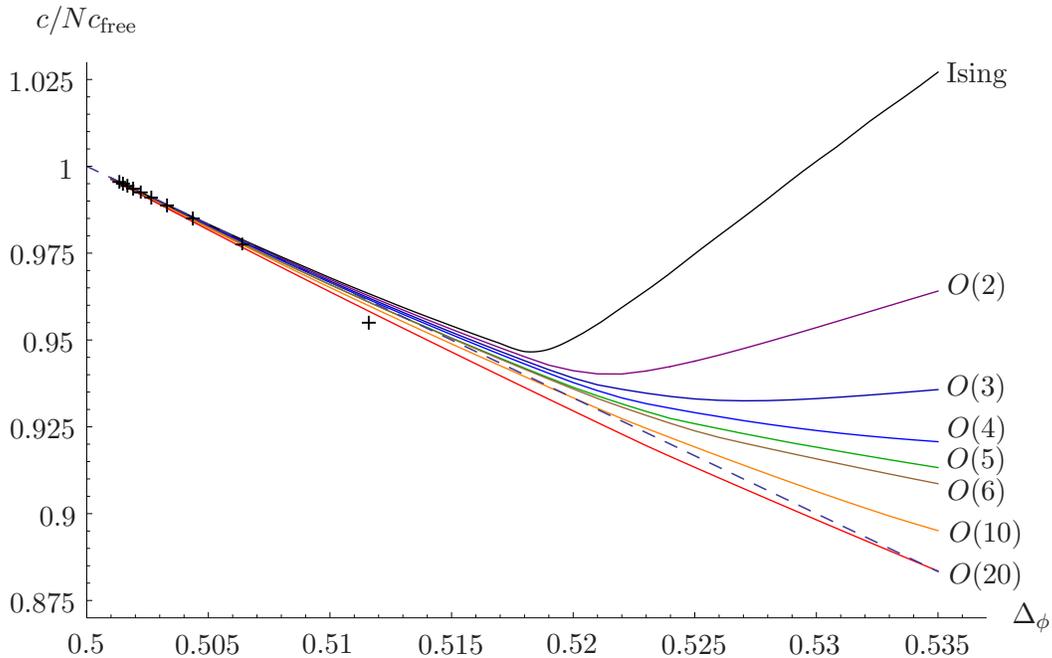}
\end{psfrags}
\caption{Lower bounds on the central charge for theories containing a scalar $\f$ transforming as a vector under $O(N)$.  We additionally assume that $\De_S,\Delta_T\geq 1$.  The black crosses show the predictions in Eq.~(\ref{eq:largeNdimensions}) from the large-$N$ expansion for $N=10,20,..., 100$.  The dashed line shows the asymptotic behavior of the central charge as a function of $\De_\f$ as $N\to\oo$.}
\label{fig:centralcharge}
\end{center}
\end{figure}

The most general bound would be obtained by making no assumptions about the operator spectrum, except that they obey unitarity conditions. However, we can obtain a somewhat stronger bound by making additional assumptions about the spectrum. In particular, we can assume there are gaps in the spectrum of singlet and symmetric tensor operators, as long as they are consistent with the results of previous subsections. Here we will assume mild gaps, $\Delta_S \ge 1$ and $\Delta_T \ge 1$. This assumption on the operator dimension spectrum is not too stringent; for example,  we know from previous determinations that $O(N)$ vector models satisfy these conditions, see Table~\ref{table:prev}.

\begin{figure}[h!]
\begin{center}
\begin{psfrags}
\def\PFGstripminus-#1{#1}%
\def\PFGshift(#1,#2)#3{\raisebox{#2}[\height][\depth]{\hbox{%
  \ifdim#1<0pt\kern#1 #3\kern\PFGstripminus#1\else\kern#1 #3\kern-#1\fi}}}%
\providecommand{\PFGstyle}{\small}%
%
\psfrag{cOverCFree}[bc][bc]{\PFGstyle $c/Nc_{\text{free}}$}%
\psfrag{deltaPhi}[cl][cl]{\PFGstyle $\Delta_\phi$}%
\psfrag{Ising}[cc][cc]{\PFGstyle $\quad\text{Ising}$}%
\psfrag{O10}[cc][cc]{\PFGstyle $\qquad O(10)$}%
\psfrag{O20}[cc][cc]{\PFGstyle $O(20)$}%
\psfrag{O2}[cc][cc]{\PFGstyle $\quad O(2)$}%
\psfrag{O3}[cc][cc]{\PFGstyle $\quad O(3)$}%
\psfrag{O4}[cc][cc]{\PFGstyle $\quad O(4)$}%
\psfrag{O5}[cc][cc]{\PFGstyle $\quad O(5)$}%
\psfrag{O6}[cc][cc]{\PFGstyle $O(6)$}%
\psfrag{ONCentralC}[bc][bc]{$O(N) ~\text{Central Charge Bounds With Singlet Gap}$}%
\psfrag{x0}[tc][tc]{\PFGstyle $0$}%
\psfrag{x11}[tc][tc]{\PFGstyle $1$}%
\psfrag{x25}[tc][tc]{\PFGstyle $0.25$}%
\psfrag{x505}[tc][tc]{\PFGstyle $0.505$}%
\psfrag{x515}[tc][tc]{\PFGstyle $0.515$}%
\psfrag{x51}[tc][tc]{\PFGstyle $0.51$}%
\psfrag{x525}[tc][tc]{\PFGstyle $0.525$}%
\psfrag{x52}[tc][tc]{\PFGstyle $0.52$}%
\psfrag{x535}[tc][tc]{\PFGstyle $0.535$}%
\psfrag{x53}[tc][tc]{\PFGstyle $0.53$}%
\psfrag{x5}[tc][tc]{\PFGstyle $0.5$}%
\psfrag{x75}[tc][tc]{\PFGstyle $0.75$}%
\psfrag{xm11}[tc][tc]{\PFGstyle $-1$}%
\psfrag{xm25}[tc][tc]{\PFGstyle $-0.25$}%
\psfrag{xm5}[tc][tc]{\PFGstyle $-0.5$}%
\psfrag{xm75}[tc][tc]{\PFGstyle $-0.75$}%
\psfrag{y0}[cr][cr]{\PFGstyle $0$}%
\psfrag{y10251}[cr][cr]{\PFGstyle $1.025$}%
\psfrag{y11}[cr][cr]{\PFGstyle $1$}%
\psfrag{y25}[cr][cr]{\PFGstyle $0.25$}%
\psfrag{y5}[cr][cr]{\PFGstyle $0.5$}%
\psfrag{y75}[cr][cr]{\PFGstyle $0.75$}%
\psfrag{y875}[cr][cr]{\PFGstyle $0.875$}%
\psfrag{y925}[cr][cr]{\PFGstyle $0.925$}%
\psfrag{y95}[cr][cr]{\PFGstyle $0.95$}%
\psfrag{y975}[cr][cr]{\PFGstyle $0.975$}%
\psfrag{y9}[cr][cr]{\PFGstyle $0.9$}%
\psfrag{ym11}[cr][cr]{\PFGstyle $-1$}%
\psfrag{ym25}[cr][cr]{\PFGstyle $-0.25$}%
\psfrag{ym5}[cr][cr]{\PFGstyle $-0.5$}%
\psfrag{ym75}[cr][cr]{\PFGstyle $-0.75$}%
\includegraphics[width=0.9\textwidth]{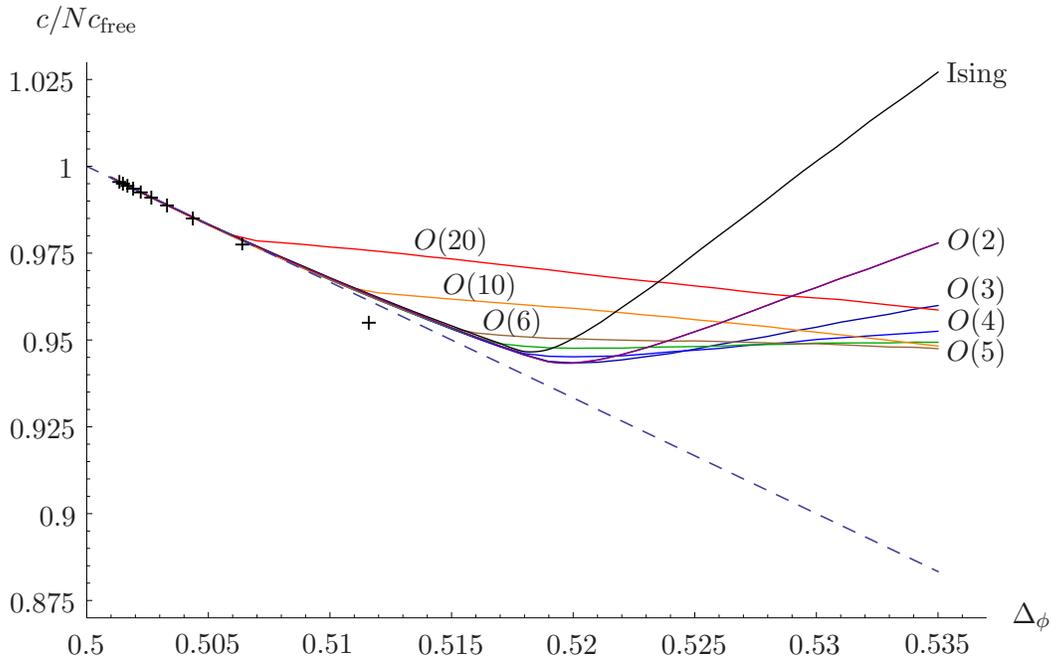}
\end{psfrags}
\caption{Lower bounds on the central charge for theories containing a scalar $\f$ transforming as a vector under $O(N)$.  In this figure we assume that the lowest dimension singlet scalar operators saturate the bounds we found in subsection~\ref{sec:boundsonsinglets}, while the symmetric tensor scalar operators are assumed to have dimensions $\Delta_T \ge 1$.  The black crosses show the predictions in Eq.~(\ref{eq:largeNdimensions}) from the large-$N$ expansion for $N=10,20,..., 100$.  The dashed line shows the asymptotic behavior of the central charge as a function of $\De_\f$ as $N\to\oo$.}
\label{fig:centralchargeS}
\end{center}
\end{figure}

The central charge bound as a function of the scalar dimension $\Delta_\phi$ is shown in Fig.~\ref{fig:centralcharge}. The central charge approximately scales linearly with $N$ (exactly in the non-interacting theory), so we have plotted $c$ scaled to $N c_\text{free}$. At low values of $\Delta_\phi$, all of the bounds approach the same asymptote. The slope of the asymptote is $-10/3$, which is the same curve that one obtains in the $N \to \infty$ limit from Eqs.~(\ref{eq:largeNdimensions}) and (\ref{eq:largeNcentralcharge}); i.e. the $O(N)$ vector model points will lie on that line for large values of $N$. 

To obtain stronger bounds on the central charge we can introduce larger gaps in the operator spectrum. In the plots of Fig.~\ref{fig:centralchargeS} we assumed that the gap in the singlet scalar spectrum saturates the bound obtained in subsection~\ref{sec:boundsonsinglets}, while the gap in the symmetric tensor scalar spectrum is kept at $\Delta_T\geq1$.  At low values of $\Delta_\phi$, the bounds again approach the same asymptote and in general don't differ too much from the bounds in the Fig.~\ref{fig:centralcharge}. However, here the bounds exhibit a change in the slope at certain value of $\Delta_\phi$. At larger values of $\Delta_\phi$ the bounds are much stronger than the ones in Fig.~\ref{fig:centralcharge}. For large $N$ values the change in the slope occurs at the $O(N)$ vector model points. At smaller $N$ the change in the slope is more gradual, but still occurs at $\Delta_\phi$ close to the known values in the $O(N)$ models.

Just like the operator dimensions in the previous subsections, the central charge values obtained using $1/N$ expansion in the case of large $N$ lie very close to the boundary line. Encouraged by these facts, we can conjecture that the central charge values will lie on the boundary even for low values of $N$. Using the values of $\Delta_\phi$ determined by other methods, we can then make a prediction for the values of $c$. These are shown, along with our bootstrap predictions for $\Delta_S$ and $\Delta_T$, in Table~\ref{table:pred}. In calculating these values of $c$ we used the bounds of Fig.~\ref{fig:centralchargeS}, since these are our strongest bounds for the $O(N)$ vector models.

\renewcommand\arraystretch{1.5}
\begin{table}
\centering
\begin{tabular}{ c c c c c } \hline\hline
\hspace{0.7cm} $N$ \hspace{0.7cm} &\hspace{0.8cm} $\Delta_\phi$ \hspace{0.8cm} &\hspace{0.8cm} $\Delta_S$ \hspace{0.8cm} &\hspace{0.8cm} $\Delta_T$ \hspace{0.8cm} & \hspace{0.8cm} $c/Nc_{\text{free}}$ \hspace{0.8cm} \\ \hline
1 & 0.51813(5) & $1.4119^{+0.0005}_{-0.0015} $&  --  &  $0.946600^{+0.000022}_{-0.000015}$  \\
2 & 0.51905(10) & $1.5118^{+0.0012}_{-0.0022}$ & $1.23613^{+0.00058}_{-0.00158}$ & $0.94365^{+0.00013}_{-0.00010}$ \\
3 & 0.51875(25) & $1.5942^{+0.0037}_{-0.0047}$ & $1.2089^{+0.0013}_{-0.0023}$ & $0.94418^{+0.00043}_{-0.00036}$ \\
4 & 0.51825(50) & $1.6674^{+0.0077}_{-0.0087}$ & $1.1864^{+0.0024}_{-0.0034}$ & $0.94581^{+0.00071}_{-0.00039}$  \\
5 & 0.5155(15) & $1.682^{+0.047}_{-0.048}$ & $1.1568^{+0.009}_{-0.010}$ & $0.9520^{+0.0040}_{-0.0030}$  \\
6 & 0.5145(15) & $1.725^{+0.052}_{-0.053}$ & $1.1401^{+0.0085}_{-0.0095}$ & $0.9547^{+0.0041}_{-0.0027}$  \\
10 & 0.51160 & $1.8690^{+0.000}_{-0.001}$ & $1.1003^{+0.000}_{-0.001}$ & 0.96394  \\
20 & 0.50639 & $1.9408^{+0.000}_{-0.001}$ & $1.0687^{+0.000}_{-0.001}$ & 0.97936 \\ \hline\hline
\end{tabular}
\caption{The values of the scalar and symmetric tensor operator dimensions and the values of the central charge saturating the obtained bound for the $O(N)$ vector model values of $\Delta_\phi$. For $N=1,2,3,4,5,6$, the value of $\Delta_{\phi}$ is taken from Table~\ref{table:prev}; for $N=10,20$ the value of $\Delta_{\phi}$ is the 3-loop large-$N$ result. The errors reflect the uncertainty in the value of $\Delta_{\phi}$. In the determinations of $\Delta_{S,T}$ we have also included a contribution to the error due to our bisection precision of $0.001$. This uncertainty is only in one direction, since the upper bound is rigorous.}
\label{table:pred}
\end{table}

\section{Discussion}
\label{sec:discussion}

Let us take a moment to reflect on the results of the previous section. First, we have discovered the remarkable fact that operator dimensions in the critical $O(N)$ vector models take on values which saturate general constraints from crossing symmetry and unitarity. This gives us an organizing principle by which we can understand why these theories are special -- gaps in the spectrum of operator dimensions are {\it maximized} in a way that is consistent with unitarity. It will be interesting to verify that this trend continues for operators of higher dimension. A promising approach to extracting more of the spectrum in these theories is to consider the locations of the zeros of $\alpha(V_{\cO})$ along the boundary~\cite{Poland:2010wg,ElShowk:2012hu,ElShowk:future}. We hope to develop this approach in the $O(N)$ models in future work. 

As far as we are aware, we have presented the first predictions for the central charge in the $O(N)$ models at small values of $N$. In doing so we have verified that there is approximately linear growth with $N$ and that $c < Nc_{\text{free}}$ for each value of $N$. It will be interesting if these predictions can be verified in lattice simulations of the $O(N)$ models -- this will require a robust lattice construction of the stress-energy tensor, which is a worthwhile task in its own right. One can also easily extend these methods to determine the flavor central charges, appearing in $\<J^{\mu} J^{\nu}\> \propto \tau$, where $J^{\mu}$ is the $O(N)$ current. 

In this work we only considered the constraints from crossing symmetry of $\<\phi_i \phi_j \phi_k \phi_l\>$. It is very interesting to extend this analysis to include constraints from other correlators, such as $\<\phi_i \phi_j \phi^2 \phi^2\>$, $\<\phi^2 \phi^2 \phi^2 \phi^2\>$, $\<\phi_i \phi_j J^{\mu} J^{\nu}\>$, or $\<\phi_i \phi_j T^{\mu\nu} T^{\rho\sigma}\>$. Such extensions are e.g. necessary in order to learn about the $\mathbb{Z}_2$ odd operators in the spectrum of these theories. Studying these correlators may also help to give a sharper criterion that can be used to determine the value of $\Delta_{\phi}$ in the $O(N)$ models, going beyond the fact that it appears to take a value near (somewhat smooth) changes in slope of the general bounds.

The recursion representation for the conformal blocks presented in section~\ref{sec:rational} is a powerful and efficient method of computing conformal blocks in any number of dimensions. This representation can for example be utilized in studies of CFTs that interpolate between $2 < D < 4$, in $D=5,7$ where conformal blocks are similarly complicated, or perhaps in constructing an argument (extending~\cite{Fitzpatrick:2013sya}) that nontrivial CFTs in large $D$ do not exist. The representation as a sum over poles in $\Delta$ may also be useful for making general analytic arguments (going beyond the large spin arguments of~\cite{Fitzpatrick:2012yx,Komargodski:2012ek}) in the context of the conformal bootstrap. The sum over poles may also be particularly interesting from the perspective of Mellin amplitudes~\cite{Mack:2009mi,Mack:2009gy,Penedones:2010ue,Fitzpatrick:2011ia,Paulos:2011ie,Fitzpatrick:2011hu,Fitzpatrick:2011dm,Paulos:2012nu,Fitzpatrick:2012cg}.

Finally, we'd like to emphasize that the $O(N)$ models at $N=2,3$ have numerous beautiful realizations in experimental condensed matter systems. E.g., the $O(2)$ model describes the superfluid transition in $^{4}$He and the bicritical point in uniaxial magnets such as GdAlO${}_3$, while the $O(3)$ model describes the Curie transition in simple isotropic magnets such as Ni, Fe, and EuO. Many more examples can be found in~\cite{Pelissetto:2000ek}. Thus, the conformal bootstrap in 3D allows one to realize the physicists' dream -- it makes quantitative predictions in strongly-interacting systems that can be experimentally tested!

\section*{Acknowledgements}
We are grateful to Rich Brower, Sheer El-Showk, George Fleming, Liam Fitzpatrick, Fred Hucht, Jared Kaplan, Miguel Paulos, Jo\~ao Penedones, Slava Rychkov, Leonardo Rastelli, Balt van Rees, Alessandro Vichi, and Sasha Zhiboedov for discussions. We would also like to thank the other organizers and participants in the Back to the Bootstrap 3 conference at CERN. The work of DSD is supported by DOE grant number DE-SC0009988.  DSD would like to thank SLAC for hospitality while this work was completed.  DP would like to thank the Galileo Galilei Institute for Theoretical Physics and the INFN for hospitality and partial support during the completion of this work. The computations in this paper were run on the Bulldog computing clusters supported by the facilities and staff of the Yale University Faculty of Arts and Sciences High Performance Computing Center, as well as the Aurora computing cluster supported by the School of Natural Sciences Computing Staff at the Institute for Advanced Study.
\newpage
\appendix

\section{Improving Rational Approximations}
\label{app:truncatingrationalapproximations}

When truncated to a finite number of poles $\De_i$, the recursion relation
\be
h_{\De,\ell}(r,\eta) &=& h^{(\oo)}_\ell(r,\eta) + \sum_i \frac{c_i r^{n_i}}{\De-\De_i}h_{\De_i+n_i,\ell_i}(r,\eta)
\ee
gives a rational approximation for $h_{\De,\ell}=r^{-\De}g_{\De,\ell}$ as a function of $\De$.  The precision of this approximation
increases as we include more and more poles $\De_i$.  However, the degree increases as well, and this can be problematic for computation.  Larger degree polynomials slow down semidefinite program solvers.

A useful compromise is to keep $n$ poles $\De_1,\dots,\De_n$ with the largest residues, and use these as a basis to approximate other poles $\De_j$ with smaller residues.  That is, for poles with small residues, we write
\be
\label{eq:shiftingpoles}
\frac{1}{\De-\De_j} &\approx& \sum_{i=1}^n \frac{a_i}{\De-\De_i}.
\ee
where the coefficients $a_i$ are chosen to make the approximation as good as possible.  In this way, we can approximately include the contribution of the pole at $\De=\De_j$ without increasing the degree of our rational function.  Note that the $\De_i$ lie below the unitarity bound $\De_\mathrm{unitarity}$, so $\De$ itself never approaches a pole when we compute CFT bounds.

How should we choose the coefficients $a_i$?  We need Eq.~(\ref{eq:shiftingpoles}) to hold to high accuracy across all $\De\ge \De_\mathrm{unitarity}$ (away from the singularities on both sides).  A method that works well in practice is to ensure that Eq.~(\ref{eq:shiftingpoles}) and its first $n/2$ derivatives hold exactly at $\De=\De_\mathrm{unitarity}$ and at $\De=\oo$.  These conditions give $n$ linear equations which determine the $a_i$.

In practice, including poles with residues less than $10^{-2}$ yields rational approximations to conformal blocks which are correct to within $10^{-9}$.  Including poles with residues less than $10^{-10}$ yields approximations correct to within $10^{-22}$.

\section{Implementation in \texttt{Mathematica} and \texttt{SDPA-GMP}}
\label{app:semidefinite}

A brief summary of our implementation is as follows.  All steps but the last are performed in \texttt{Mathematica}.

\begin{enumerate}
\item We compute a rational approximation for derivatives of conformal blocks at the crossing symmetric point $\ptl_r^m \ptl_\eta^n g_{\De,\ell}\approx r^{\De}P^{(m,n)}_{\ell}(\De)/Q_\ell(\De)$, where $r=3-2\sqrt{2}$.  This can be done using either the Gegenbauer expansion and Casimir equation described in \cite{Hogervorst:2013sma}, or more efficiently using our recursion relation~(\ref{eq:recursionh}).  The recursion relation can be implemented numerically in the space of vectors of $(r,\eta)$ derivatives, where multiplication by $r^k$ is a matrix on this space.  We compute approximations up to order $60$ in the $r$-expansion.

\item We approximate ``small" poles as described in Appendix~\ref{app:truncatingrationalapproximations}, resulting in a new rational approximation with smaller degree $\ptl_r^m \ptl_\eta^n g_{\De,\ell}\approx r^{\De}p^{(m,n)}_{\ell}(\De)/q_\ell(\De)$.  To decide which poles to keep and which poles to approximate, we choose a threshold value $\th$ and compute
\be
\th_i &\equiv& \max_{m,n} \left|\frac{
    \mathop{\mathrm{Res}}_{\De\to\De_i} \ptl_r^m \ptl_\eta^n g_{\De,\ell}
}{
    \ptl_r^m \ptl_\eta^n g_{\De,\ell}|_{\De=\textrm{unitarity bound}}
    }\right| .
\ee 
Poles with $\th_i\leq\th$ are approximated in terms of other poles.  In practice, we found that $\th=10^{-2}$ gives a good tradeoff between accuracy and speed.  We have checked that our results remain essentially unchanged as $\th$ is varied between $10^{-2}$ and $10^{-12}$.  After reducing the degree of our rational approximation, the factors $r^\De/q_\ell(\De)$ can be discarded.  Henceforth, we will use ``$\sim$" to indicate approximate equality up to an overall positive function of $\De$.

\item To compute a bound, we need derivatives $\ptl_z^m \ptl_{\bar z}^n V_{R,\De,\ell}$, where the vectors $V_{R,\De,\ell}$ are defined in~(\ref{eq:vectordefinitions}).  These are linearly related to the vectors of derivatives of $g_{\De,\ell}$,
\be
\ptl_z^m \ptl_{\bar z}^n V_{R,\De,\ell,i} &\sim& (M_{R,\De_\f})^{mn}_i{}_{rs} P^{(r,s)}_{\ell}(\De)
\ee
where $i=1,2,3$ runs over the components of $V_{R,\De,\ell}$ and $M_{R,\De_\f}$ is a matrix depending only on the representation/channel $R$ and the external operator dimension $\De_\f$.

\item In section~\ref{sec:formulation}, we defined a semidefinite program as an affine optimization problem which can include constraints of the form
\be
\label{eq:polynomialsemidefinite}
\a(P_i(x))\textrm{ for all $x\geq0$, where $P_i(x)$ are polynomials in $x$.}
\ee
More precisely, semidefinite programs can include matrix inequalities of the form
\be
\label{eq:matrixsemidefinite}
X \textrm{ is positive semidefinite},
\ee
where $X$ is a matrix of variables (which might be subject to additional linear constraints).  The transformation of a set of inequalities from the form~(\ref{eq:polynomialsemidefinite}) to the form~(\ref{eq:matrixsemidefinite}) is standard in the optimization literature~\cite{Vandenberghe:1996} and is described in detail in \cite{Poland:2011ey}.  
We transform our polynomial inequalities
\be
a_{mn}^i(M_{R,\De_\f})^{mn}_i{}_{rs}P_\ell^{(r,s)}(\De_{\min,\ell}+x) \geq 0\quad\textrm{for $x\geq 0$}
\ee
into matrix inequalities in this fashion.

\item Once written in terms of matrix inequalities, our semidefinite program can be solved using a variety of freely available tools.  For this work, we use the solver \texttt{SDPA-GMP} with the parameters listed in Table~\ref{tab:SDPAparams}.   Our plots are computed in parallel by assigning each point to an individual cluster node.  Our cluster management software is written in Cloud Haskell.

\begin{table}
\begin{center}
\begin{tabular}{r | l}
parameter & value \\
\hline
maxIteration & $1000$ \\
epsilonStar & $10^{-20}$ ($10^{-10}$) \\
lambdaStar & $10^{20}$ \\
omegaStar & $10^{20}$ \\
lowerBound & $-10^{40}$ \\
upperBound & $10^{40}$ \\
betaStar & $0.1$ \\
betaBar & $0.3$ \\
gammaStar & $0.7$ \\
epsilonDash & $10^{-20}$ ($10^{-10}$) \\
precision & $200$ ($300$)
\end{tabular}
\end{center}
\caption{\texttt{SDPA-GMP} parameters used in the calculation of the operator dimension bounds. In parentheses are the values of the parameters used in the central charge bounds.}
\label{tab:SDPAparams}
\end{table}

\end{enumerate}

\bibliography{Biblio}{}
\bibliographystyle{utphys}

\end{document}